\documentclass[aps,showpacs,nofootinbib,Arial]{revtex4-2}
\usepackage{graphicx}
\usepackage{amssymb}
\usepackage{amsmath}
\usepackage{graphics}
\usepackage{dcolumn}
\usepackage{bm}
\usepackage{color}
\usepackage{epstopdf}
\usepackage{lipsum}
\usepackage{hyperref} 
\usepackage{ulem}

\begin{document}
	
	\newcommand{\blue}[1]{\textcolor{blue}{#1}}
	\newcommand{\new}{\blue}
	\newcommand{\red}[1]{\textcolor{red}{#1}}
	\newcommand{\att}{\red}
	
\title{Entropy Production in the Inflationary Epoch Using the Gouy-Stodola Theorem}

\author{R. H. Longaresi\footnote{longaresi@ufscar.br}}
	
\address{Federal University of S\~ao Carlos, Sorocaba CEP 18052780, Brazil}
	
\author{S. D.  Campos\footnote{sergiodc@ufscar.br}}
	
\address{Applied Mathematics Laboratory-CCTS/DFQM,\\
Federal University of S\~ao Carlos, Sorocaba CEP 18052780, Brazil}
	
\begin{abstract}
In this work, we use the Gouy-Stodola theorem to calculate the entropy production rate in the inflationary epoch of the universe. This theorem allows us the simple calculation of entropy and entropy production rate occasioned by the decaying of the inflaton scalar field.  Both the entropy and entropy production rate achieve large values, agreeing with the expected values present in the literature.
\end{abstract}
	
\date{\today}
\maketitle
	
\section{Introduction}\label{sec:int}
	
The laws of thermodynamics are one of the cornerstones of physics, and the second law possesses the same importance as the energy conservation law. Nonetheless, correctly defining entropy is a difficult task because it depends strongly on the physical properties of the system under study. For example, the entropy can be related to one-quarter of the black hole event horizon area \cite{hawking,bekenstein}, or Shannon's information disorder approach \cite{shannon,shannonweaver}, or energy dissipation in the classic thermodynamics approach. Moreover, the correct definition of entropy is a problem that permeates several issues, for example, the study of entropy production in rotating black holes \cite{cvetic}, in the so-called Ekpyrotic universe \cite{li}, in Szekeres spacetimes \cite{herrera}, quark-gluon plasma \cite{mattiello}, light clusters \cite{vermani}, and decoherence processes \cite{fries}, among others.
	
In classical mechanics, the existence of non-conservative forces interferes with entropy production, resulting in they should be taken into account for the correct system description. According to literature, there are different ways to include entropy calculation in classical mechanics, such as the canonical Hamiltonian or Lagrangian formalism. These formalism allows calculating the rate of entropy production \cite{J.Silverberg.A.Widom.Amer.J.Phys.75.993.2007} as well as considering dissipative processes \cite{A.N.Kaufman.Phys.Lett.A100.419.1983} or the method of molecular dynamics by modifying the Newtonian dynamics \cite{S.Nose.Mol.Phys.52.255.1984,W.G.Hoover.Phys.Rev.A.31.1695.1985}. 

In mechanical systems, entropy generation can be computed through the Silverberg and Widom approach \cite{J.Silverberg.A.Widom.Amer.J.Phys.75.993.2007} or by the Gouy-Stodola theorem \cite{M.Gouy.J.de.Phys.8.35.1889,A.Stodola.Zeitschr.d.Verein.deutscher.Ingenieure.32.1086.1898,A.Bejan.G.Tsatsaronis.M.Moran.Thermal.Design.and.Optimization.JWS.1996}, for instance. This theorem is a less known result even in engineering, where its formal applicability conditions are not clearly stated. However, its original formulation is mathematically quite simple: The work lost in an open system is directly proportional to the amount of entropy generated in the process \cite{R.Pal.Int.J.Mech.Eng.Ed.45.2.142.2017}. The work lost by the system, in its turn, is given by the difference between the work done by the reversible (free of damping forces) and the irreversible processes, where different kinds of drag forces can be introduced. It should be noted this theorem hold even for adiabatic systems \cite{R.Pal.Int.J.Mech.Eng.Ed.45.2.142.2017}.
	
In the present paper, we first introduce the rate of entropy production given by the Gouy-Stodola theorem by considering two different systems. The first system is a simple pendulum subject to a damping force. After that, we modify the system introducing the possibility of the parametric resonance mechanism but still considering the same damping force. In both mechanical conditions, one obtains the entropy produced as stated by the Gouy-Stodola theorem.
	
As the second and most important step, we apply the Gouy-Stodola theorem for a system taking into account parametric resonance to the particle production in the early universe. In the simple case of the scalar inflaton field $\phi$ decaying into a scalar $\chi$ particle, the entropy production for $\phi$ can be calculated using the damping term of the equation of motion for the inflaton. For the $\chi$ particles, is used the same method. As well-known, exponential particle production due to the parametric resonance mechanism can fastly populate the universe \cite{A.D.Dolgov.D.P.Kirilova.Yad.Fiz.51.273.1990.Sov.J.Nucl.Phys.51.172.1990,J.H.Traschen.R.H.Brandenberger.Phys.Rev.D42.2491.1990}. Thus, entropy production due to this effect may be an important ingredient in explaining the huge amount of entropy observed in the universe
\cite{J.Romero.M.Bellini.Nuovo Cim.B124.861-868.2009,C.A.Egan.C.H.Lineweaver.Astrophys.J.710.1825-1834.2010}. 

It is important to stress that we are not interested in the details of the particle production mechanism, which has been exhaustively studied over decades as shown in the literature \cite{A.D.Dolgov.D.P.Kirilova.Yad.Fiz.51.273.1990.Sov.J.Nucl.Phys.51.172.1990,J.H.Traschen.R.H.Brandenberger.Phys.Rev.D42.2491.1990,L.Kofman.A.Linde.A.Starobinsky.Phys.Rev.Lett.73.3195-3198.1994,Y.Shtanov.J.Traschen.R.Brandenberger.Phys.Rev.D51.5438-5455.1995,M.Yoshimura.Prog.Theor.Phys.94.873-898.1995,I.Zlatev.G.Huey.P.J.Steinhardt.Phys.Rev.D57.2152-2157.1998,V.Zanchin.A.MaiaJr.W.Craig.R.Brandenberger.Phys.Rev.D60.023505.1999,P.Channuie.P.Koad.Phys.Rev.D94.043528.2016,I.Romualdo.L.Hackl.N.Yokomizo.Phys.Rev.D100.065022.2019}. Our main goal is to show that the Gouy-Stodola theorem can be used as a simple tool, allowing the calculation of entropy production even for complex physical situations. We are interested in showing that entropy generation in the inflationary epoch depends on the considered scalar field. Furthermore, entropy and entropy production due to $\phi$ depends strongly on both its mass, $m_\phi$, and its self-coupling constant, $\lambda$.  However, for some values, entropy associated with $\phi$ can achieve $>10^{98}$. Thus, it is important to understand the role played by each scalar field in entropy generation during the inflationary era.

This paper is organized as follows. Section \ref{sec:dev} presents the simple pendulum subject to a damping force, with and without the parametric resonance mechanism, and calculates the entropy produced in both cases. In Section \ref{sec:early}, one calculates the entropy produced according to the Gouy-Stodola theorem. Section \ref{sec:fr} presents our final remarks and criticism.
	
\section{Entropy Production and the Simple Pendulum}\label{sec:dev}
For the sake of clarity, the Gouy-Stodola theorem is applied to the simple pendulum in two situations. In the first step, one considers only the simple pendulum subject of a drag force. In the second step, we introduce the parametric resonance mechanism with the same drag force as before.

\subsection{Entropy Production}

Entropy production depends on specific details of the physical system (its dynamics), and there are several frameworks developed to define entropy production in some open systems \cite{L.Onsager.Phys.Rev.38.2265.1931,S.R.deGroot.P.Mazur.Non.Equilibrium.Thermodynamics.North.Holland.Physics.1961,M.Campisi.P.Hanggi.P.Talkner.Rev.Mod.Phys.83.771.2011,M.Esposito.U.HarbolaS.Mukamel.Rev.Mod.Phys.81.1665.2009,S.Ito.M.Oizumi.S.Amari.Phys.Rev.Research.2.033048.2020,G.A.Weiderpass.A.O.Caldeira.Phys.Rev.E.102.032102.2020}.
For our purposes, it is sufficient to show how the entropy production rate is related to the work done in reversible and irreversible systems. Firstly, notice that the Clausius inequality can be written as
\begin{eqnarray}\label{eq:clausius_1}
    \oint_i^f \frac{\delta Q}{T}\leq 0,
\end{eqnarray}

\noindent where one assumes the system undergoes some process from an initial state $i$ up to the final state $f$. It is important to stress that the integral above defined is calculated on the system boundary, and for convenience, one writes
\begin{eqnarray}\label{eq:clausius_2}
    \oint_i^f \frac{\delta Q}{T}=-\Sigma
\end{eqnarray}

\noindent where one defines that $\Sigma=\Sigma(t)$ is zero for reversible processes as well as positive for irreversible processes, and $\delta Q$ is the heat transfer due to a specific process (isothermic, isobaric or isocoric), and $T$ is system temperature. For a reversible process, which ensure the equality in (\ref{eq:clausius_2}), $\Sigma$ is just the entropy of the system
\begin{eqnarray}\label{eq:clausius_3}
    S_i-S_f=\oint_f^i \frac{\delta Q}{T},
\end{eqnarray}

\noindent which is, in fact, zero since the process is reversible. Observe that the temperature which occurs the heat transfer from the reservoir to the system is constant.

Now, consider a system consisting of two cycles. The first cycle is reversible, while the second is irreversible. Then, one writes
\begin{eqnarray}\label{eq:ent_prod_1}
    \Sigma = \Delta S - \oint_f^i \frac{\delta Q}{T}\geq 0,
\end{eqnarray}

\noindent where the integral takes into account the irreversible process. From the above result, one observes that $\Sigma$ is due to the irreversible process given by the integral. Thus, one defines $\Sigma$ as the entropy production, which is always a non-negative quantity. From definition (\ref{eq:ent_prod_1}), one introduces the entropy production rate 
\begin{eqnarray}
    \dot{\Sigma}=\dot{S}-\frac{\dot{Q}}{T},
\end{eqnarray}

\noindent where the dot means the usual time derivative $d/dt$. 

Of course, energy conservation holds for reversible and irreversible systems. For general open systems composed of $n$ independent processes, one can write energy conservation rate as (following close Ref. \cite{A.Bejan.Int.J.Energy.Res.26.545.2002})
\begin{eqnarray}
    \dot E=\sum_{i=0}^n\dot{Q}_i -\dot W +c_1(m)=\dot{Q}_0+\sum_{i=1}^n\dot{Q}_i -\dot W +c_1(m),
\end{eqnarray}

\noindent where $c_1(m)$ is the sum of the contribution of the finite mass rate to the energy rate, and $W$ is the work due to the irreversible forces acting on the system. Moreover, one writes the entropy production rate for open systems
\begin{eqnarray}
    \dot\Sigma=\dot S-\sum_{i=0}^n\frac{\dot{Q}_i}{T_i}+c_2(m)=\dot S-\frac{\dot{Q}_0}{T_0} - \sum_{i=1}^n\frac{\dot{Q}_i}{T_i}+c_2(m)\geq 0,
\end{eqnarray}

\noindent where, now, $c_2(m)$ represents the sum of the contribution of the finite mass rate to the entropy production rate. The rate of the heat transfer $\dot{Q}_0$ can be eliminated by combining the above results, implying the work rate for open systems is given by
\begin{eqnarray}\label{eq:irrev}
    \dot{W}=-(\dot{E}-T_0\dot{S})+\sum_{i=1}^n\left[1-\frac{T_0}{T_i}\right]\dot{Q}_i+c_1(m)+T_0c_2(m)-T_0\dot\Sigma.
\end{eqnarray}

For reversible systems, the last term on the r.h.s of the above result is null, resulting
\begin{eqnarray}\label{eq:rev}
    \dot{W}_r=-(\dot{E}-T_0\dot{S})+\sum_{i=1}^n\left[1-\frac{T_0}{T_i}\right]\dot{Q}_i+c_1(m)+T_0c_2(m).
\end{eqnarray}


Subtracting (\ref{eq:irrev}) from (\ref{eq:rev}), one obtains the entropy production rate in terms of the difference between the work rate for the reversible process and the irreversible one
\begin{eqnarray}\label{eq:eq_sigma1}
    T_0\dot\Sigma= \dot{W}_{r}-\dot{W}.
\end{eqnarray}

The above result is known as the Gouy-Stodola theorem \cite{M.Gouy.J.de.Phys.8.35.1889,A.Stodola.Zeitschr.d.Verein.deutscher.Ingenieure.32.1086.1898,A.Bejan.G.Tsatsaronis.M.Moran.Thermal.Design.and.Optimization.JWS.1996}. In the next subsections, one discusses this result for both a simple pendulum and the one subject to the parametric resonance phenomenon.

\subsection{The Simple Pendulum}

One considers a damped pendulum, i.e. a weight of mass $m$ (bob) suspended from a pivot so that it can oscillate freely subject to the restoring force of gravity $\vec{g}$. In this example, one uses a very simple damping force given by
	\begin{eqnarray}\label{eq:drag_force}
		D(v)=-b v=-b\dot{s}(t),
	\end{eqnarray} 
	
\noindent where the motion of the bob is along the arc $s(t)=l\theta(t)$, being $l$ the inextensible string length of negligible mass. The drag parameter $b$ measures the strength of the damping force. Of course, the presence of a drag force implies the system is non-conservative (irreversible). The work $W_D$  done by the drag force $D(\theta)$ is given by
\begin{eqnarray}\label{eq:work}
    W_D =\new{-} \int_{\theta_0}^{\theta}D(\theta')ld\theta'.
\end{eqnarray}

The resulting total energy $E$ describing the system depends on $\theta$, being given by
	\begin{eqnarray}
		E=K+U+W_{D}=\frac{ml^2}{2}\dot{\theta}^2+mgl(1-\cos\theta)-bl^2\dot\theta(\theta - \theta_0),
	\end{eqnarray}
	
\noindent where $K$ is the kinetic energy, $U$ is the potential energy, and $\theta_0$ is the initial angle. 
	
	
For mechanical systems, the Gouy-Stodola theorem \cite{M.Gouy.J.de.Phys.8.35.1889,A.Stodola.Zeitschr.d.Verein.deutscher.Ingenieure.32.1086.1898} relates the entropy production to the difference between the time derivative of the reversible and irreversible work realized by the forces acting in the system. In the present case and according to Eq. (\ref{eq:eq_sigma1}), this difference is given by
	\begin{eqnarray}\label{eq:ent_prod}
		\dot{\Sigma}=	\frac{\dot{W}_r-\dot{W}}{T}=\frac{\dot{W}_r-\dot{W}_{D}}{T}=\frac{bl^2}{T}[\dot{\theta}^2-(\theta-\theta_0)\ddot{\theta}],
	\end{eqnarray}
	
\noindent where $T$ is the system temperature taken as constant. Considering clockwise oscillation as initial condition, (\ref{eq:ent_prod}) can be rewritten,
\begin{eqnarray}\label{eq:ent_prod2}
		\dot{\Sigma}=\frac{bl^2}{T}[\dot{\theta}^2+(\theta-\theta_0)\ddot{\theta}].
	\end{eqnarray}
	
It should be noted that $\dot{\Sigma}$ shown above only depends on the work done by the drag force given by equation (\ref{eq:drag_force}). Thus, entropy production is strictly related to energy dissipation. Therefore, it is important to keep in mind the difference between definitions of entropy and entropy production: Entropy can be calculated for any physical system, and it is an extensive property of the system (state function) and, therefore, it is process-independent. As the dissipative force characterizes entropy production, entropy production is process-dependent. Both entropy and entropy production exists for reversible and irreversible processes. However, in the first process both remain constant while not for the latter. Notice that if the drag force is neglected, then the entropy related to the conservative system (reversible) achieves its maximum at the minimum of the kinetic energy (the maximum entropy state) and the entropy production, accordingly to relation (\ref{eq:ent_prod2}), is zero.

The equation of motion for $\theta(t)$ can help us to understand the time-dependent behavior of $\dot{\Sigma}$. Considering the drag force, one has the well-known equation
	\begin{eqnarray}\label{eq:eq_motion_dragforce_1}
		\ddot{\theta}(t)+\kappa\dot{\theta}+\omega^2\sin\theta=0,
	\end{eqnarray}
	
\noindent where $\kappa=b/m$ and $\omega^2=g/l$. Considering the small-angle solution, $\sin\theta\approx \theta$, the simplest solution is given by
	\begin{eqnarray}\label{eq:gen_sol}
		\theta(t)=\alpha e^{-\gamma_1 t}+\beta e^{-\gamma_2 t},
	\end{eqnarray} 
	
\noindent where $\alpha$ and $\beta$ depends on the initial conditions for the system. Moreover, one defines $\gamma_1$ and $\gamma_2$ as
	\begin{eqnarray}
		-\gamma_1=-(\kappa+\sqrt{\kappa^2-\omega^2}),\\
		-\gamma_2=-(\kappa-\sqrt{\kappa^2-\omega^2}).
	\end{eqnarray}
	
Regarding the above calculations and taking into account the solution for $\theta(t)$ given by (\ref{eq:gen_sol}), one obtains the total entropy produced by the drag force in the result (\ref{eq:ent_prod}). One adopts here the SI system and, for the sake of simplicity, one sets $l=1$ m, $|\vec{g}|=10$ m/s$^2$, $m=1$ kg, and $T=298$ K. The parameter $b$ is defined according to the damped cases from classical mechanics. For $\kappa^2<\omega^2$, one has the underdamped case, while for $\kappa^2>\omega^2$ and $\kappa^2=\omega^2$ one obtains the over- and critically damped cases, respectively. It is important to point out that due to the oscillatory character of the solutions for $\theta(t)$, entropy and entropy production also oscillate, assuming positive and negative values as $t$ rises. Therefore, in all figures, we display the absolute value of entropy and entropy production.

\begin{figure}
		\centering
		\includegraphics[scale=0.32]{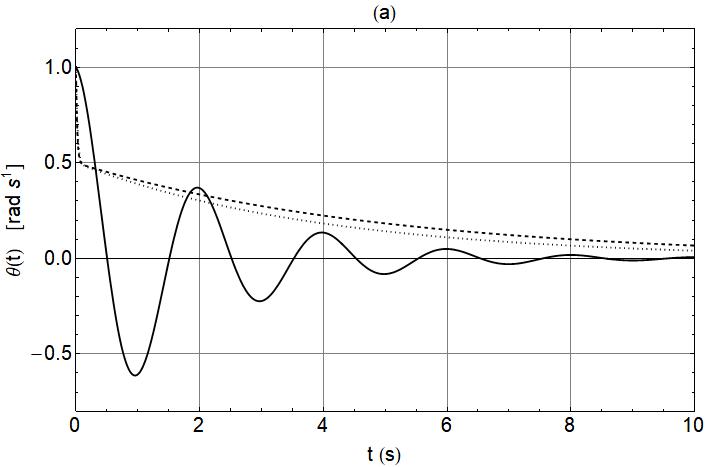}
		\includegraphics[scale=0.315]{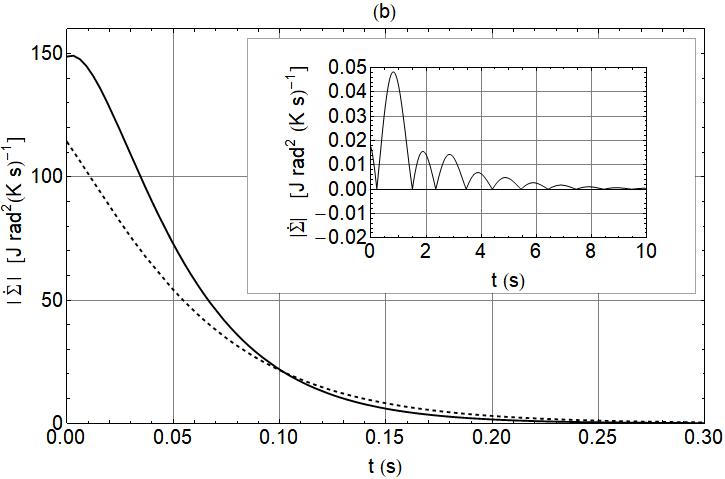}
		\caption{\label{fig:fig_drag_1}Panel (a) shows $\theta(t)$ for the simple pendulum. Solid line represents the underdamped while dashed and dotted lines describes the over- and critically damped cases, respectively. Panel (b) shows the absolute value for the entropy production rate, $\dot\Sigma$, considering the underdamped (solid line), overdamped (dashed line), and critically damped (dotted line) cases. The parameters used are displayed along the text.}
	\end{figure}

Figure $\ref{fig:fig_drag_1}$(a) shows $\theta(t)$ according to $\kappa^2 = 1$ s$^{-2}$ (solid line), and $\kappa^2 = 50$ s$^{-2}$, for dashed and dotted lines, respectively. Figure \ref{fig:fig_drag_1}(b) shows the absolute value for the entropy production until the pendulum motion ceases, reaching its equilibrium state. The inner panel in Figure \ref{fig:fig_drag_1}(b) represents the entropy production rate in the underdamped case. It should be noted that in the over- and critically damped cases shown by dashed and dotted curves in both panels, the entropy production is positive since there is no inversion in the pendulum motion, i.e. the values adopted for $\kappa$ does not allow oscillatory motion.

\begin{figure}
		\centering
		\includegraphics[scale=0.31]{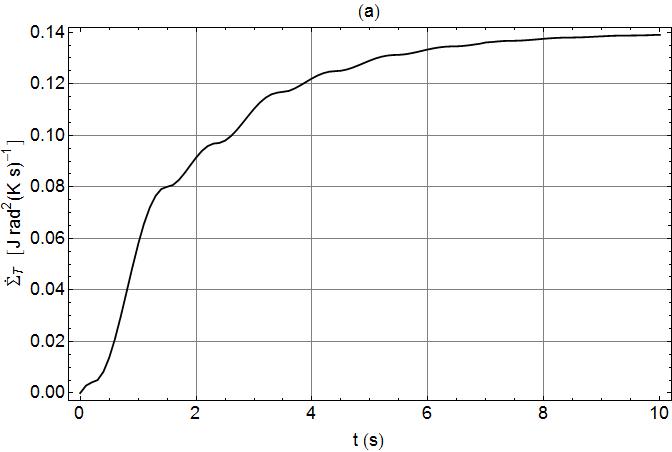}
		\includegraphics[scale=0.315]{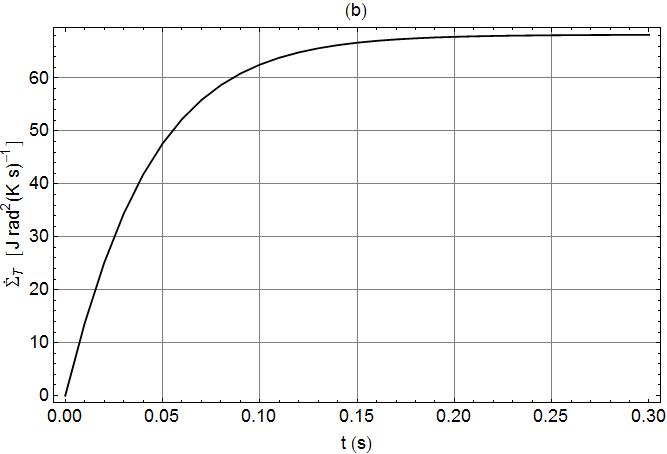}
		\caption{\label{fig:fig_accumulated}Accumulated entropy production rate for the oscillatory, panel (a), and overdamped case, panel (b).}
	\end{figure}

The sum of entropy production rates occurring at each period should be a growing function since it corresponds to the accumulated entropy production rate. Figure \ref{fig:fig_accumulated} shows the numerical calculation for the accumulated entropy production rate, $\dot{\Sigma}_T$, for the oscillatory case, panel (a), and for the overdamped case, panel (b). For both situations, the accumulated entropy production is a growing function of $t$. Also notice that since the entropy production rate tends to zero due to the damped term, the accumulated result approaches a constant value.

From Eq. (\ref{eq:ent_prod}), we can observe that the zeros of $\dot\Sigma(t)$ are given by
\begin{eqnarray}\label{eq:zeros}
    \dot\theta^2(t)=-(\theta(t)-\theta_0)\ddot\theta(t),
\end{eqnarray}

\noindent which implies the product in the r.h.s of result (\ref{eq:zeros}) is always positive for all $t$ since $\theta(t)\leq \theta_0$. The equality is valid for the simple harmonic oscillator where non-conservative forces are absent and the system can return to the same energy state ($\Delta S=0$), and the resulting entropy production variation, according to result (\ref{eq:ent_prod}), should also be zero. Therefore, the result (\ref{eq:zeros}) establishes conditions for the oscillatory motion when considering non-conservative forces described in (\ref{eq:drag_force}). We can realize that the unavailable energy for the weight force to do work, which means the dissipated energy by the drag force, is responsible for the entropy production. Due to the inversion in the pendulum motion, the entropy production rates oscillate, and the zeros of the entropy production rate correspond to the intermediates equilibrium states of the pendulum motion.
	
\subsection{The Simple Pendulum: Parametric Resonance}
	
Parametric resonance is a well-known phenomenon associated with time-dependent frequency, which is the parameter of the problem. The general case without a drag force can be written as
	\begin{eqnarray}\label{eq:mathieu}
		\ddot{f}(t)+\omega_0(t)f(t)=0,
	\end{eqnarray}
	
\noindent where, for convenience, one defines the time-dependent frequency as
	\begin{eqnarray}
		\omega_0(t)=\omega_0^2[1+\zeta \cos(\omega_1 t)],
	\end{eqnarray}
	
\noindent being $\omega_0$, $\omega_1$, and $\zeta$ real parameters. The result given by (\ref{eq:mathieu}) is the so-called Mathieu equation \cite{Landau.Lifschitz.Mechanics} commonly used to explain particle production processes in cosmology \cite{A.D.Dolgov.D.P.Kirilova.Yad.Fiz.51.273.1990.Sov.J.Nucl.Phys.51.172.1990,J.H.Traschen.R.H.Brandenberger.Phys.Rev.D42.2491.1990}. In general, the amplitude of the solution depends on the relation between $\omega_1$ and $\omega_0$. When $\omega_1\approx 2\omega_0$, the amplitude of the oscillations grows exponentially with time, leading to an explosive particle production. It is convenient to write
	\begin{eqnarray}
		\omega_1=2\omega_0+\epsilon,
	\end{eqnarray}
	
\noindent where $\epsilon<\!\!<1$ is a small perturbation. Parametric resonance condition is achieved if \cite{Landau.Lifschitz.Mechanics}
	\begin{eqnarray}
		\epsilon<\left| \frac{\zeta\omega_0}{2} \right|.
	\end{eqnarray}
	
Considering the presence of a drag force, one has the modified Mathieu equation for the simple pendulum written as
	\begin{eqnarray}\label{eq:eq_motion_dragforce_2}
		\ddot{\theta}(t)+\kappa\dot{\theta}+\omega_0(t)\sin\theta=0,
	\end{eqnarray}
	
\noindent where one uses
	\begin{eqnarray}
		\omega_0(t)=\omega_0^2\left[\frac{g}{l}+\zeta \cos(2\omega_0+\epsilon) t\right].
	\end{eqnarray}
	
Of course, for $\zeta=0$, one recovers the non-parametric resonance case taking $\omega_0=1$. Considering $\sin \theta \approx \theta$, one writes
    \begin{eqnarray}
	\label{eq:eq_motion_dragforce_3}
		\ddot{\theta}(t)+\kappa\dot{\theta}(t)+\omega_0(t) \theta(t)=0.
	\end{eqnarray}

For the sake of simplicity, one writes a possible solution for the equation (\ref{eq:eq_motion_dragforce_3}) as  
\begin{eqnarray}\label{eq:solution_1}
    \theta(t)\approx (\cos(\omega_0+\epsilon/2)t + \sin(\omega_0+\epsilon/2)t)e^{-\kappa t}.
\end{eqnarray}

It should be stressed the proper solution of equation (\ref{eq:eq_motion_dragforce_3}) encompasses Mathieu functions, defined through a combination of sine and cosine functions. However, for our purposes in this section, the ansatz (\ref{eq:solution_1}) is sufficient to show the entropy production in the parametric resonance case.

\begin{figure}
		\centering
		\includegraphics[scale=0.315]{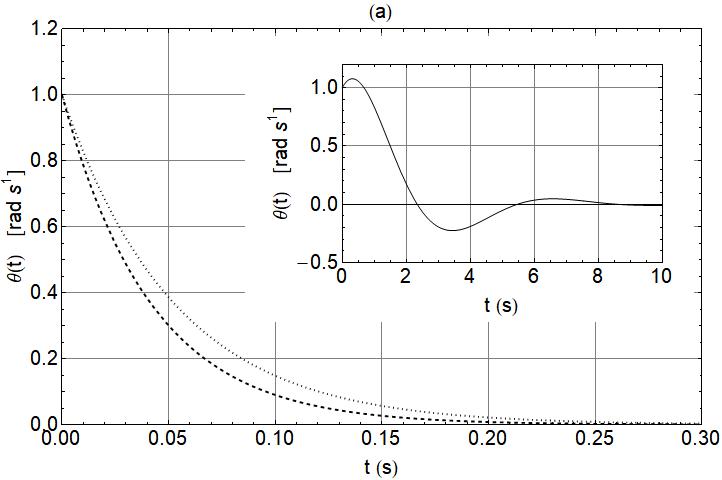}
		\includegraphics[scale=0.32]{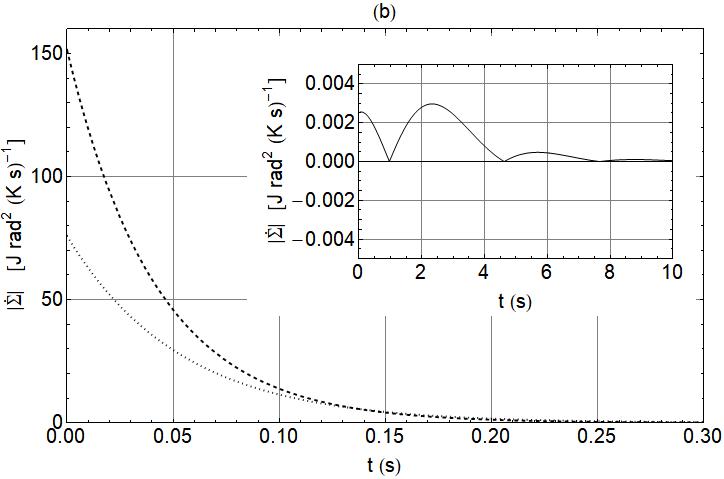}
		\caption{\label{fig:fig_drag_2}Results taking into account the parametric resonance for $\theta(t)$ (panel a), and the absolute value of $\dot{\Sigma}$ (panel b).}
	\end{figure}

Figure \ref{fig:fig_drag_2} displays the results for $\theta(t)$ and the absolute value of $\dot{\Sigma}(t)$ according to solution (\ref{eq:solution_1}). As expected, the parametric resonance case is similar to the result given by the damped pendulum.

\section{Particle Production in the Inflationary Era}\label{sec:early}

One assumes the inflationary period can provide a causal mechanism for generating several structures observed in the present-day Universe, such as the primordial perturbations responsible for the formation of galaxies \cite{K.H.Seleim.A.A.El-Zant.A.M.Abdel-Moneim.Phys.Rev.D102.063505.2020} and the homogeneity problem \cite{V.Pervushin.D.Proskurin.Grav.Cosmol.Suppl.8N1.161-167.2002}. The inflaton scalar field $\phi$ is the leading contribution to the energy-momentum tensor possessing here the following Lagrangian
\begin{eqnarray}\label{eq:eq_lagrangeana}
  \mathcal{L}(\phi)=\frac{1}{2}(\partial_\mu \phi)^2 - V(\phi),
\end{eqnarray}

\noindent where $V(\phi)$ is the effective potential of the scalar field $\phi$. The choice of such potential depends on the interactions we want to assume. From now on, one adopts the natural unit system where $c=h=k_B=1$.

The Einstein field equations control the evolution of the flat Friedman-Robertson-Walker universe,
\begin{eqnarray}
   \left(\frac{\dot{a}}{a} \right) H^2=\frac{8\pi}{3M_p^2}\left(\frac{1}{2}\dot{\phi}^2+V(\phi) \right),
\end{eqnarray}

\noindent where $a=a(t)$ is the scale factor, $H=\dot{a}(t)/a(t)$ is the Hubble parameter, and $M_p$ is Planck mass. The Klein-Gordon equation for $\phi$ depends on the choice of $V(\phi)$ and can be written as
\begin{eqnarray}\label{eq:kg_1}
\ddot{\phi}+3H\dot{\phi}+V'(\phi)=0,
\end{eqnarray}

\noindent where $V'(\phi)=dV(\phi)/d\phi$ and $3H\dot{\phi}$ act as a "friction term". In general, the interest resides on the decay of $\phi$ into scalar $\chi$ particles, avoiding the Pauli Exclusion Principle. Then, it is necessary to know the total decay rate, and for the case where $\phi$ decays into two scalar particles, one writes \cite{A.D.Linde.Particle.Physics.and.Inflationary.Cosmology.Harwood.Chur.Switzerland.1990}
\begin{eqnarray}
   \Gamma=\Gamma (\phi\rightarrow\chi\chi)=\frac{h^4\sigma^2}{8m_{\phi}},
\end{eqnarray}

\noindent where $h$ is coupling constant and $\sigma$ is a constant with dimension of mass. Of course, we are supposing the mass of $\phi$ scalar field $m_\phi> m_\chi$, where $m_\chi$ is mass of $\chi$ particles. Moreover, during the oscillating period, one also suppose $m_\phi>\!\!>H$. Thus, one adds the term $\Gamma\dot{\phi}$ to equation (\ref{eq:kg_1}), representing the back reaction of produced particles. The decay rate act as a friction term, resulting in a phenomenological description for $\phi$, written as
\begin{eqnarray}\label{eq:kg_2}
    \ddot{\phi}+(3H+\Gamma)\dot{\phi}+V'(\phi)=0.
\end{eqnarray}

The effective potential $V(\phi)$, chosen here by its simplicity, is given here by the usual $\lambda\phi^4$
\begin{eqnarray}
V(\phi)=\frac{m_\phi^2}{2}\phi^2-\frac{\lambda}{4}\phi^4,
\end{eqnarray}

\noindent where $\lambda$ is the $\phi$ self-coupling constant and should be chosen in order to keep the flatness of $V(\phi)$. Of course, there are several potential to $\phi$, each one depending on the approach performed \cite{J.Martin.C.Ringeval.R.Trotta.V.Vennin.JCAP.1403.039.2014}. The value of $\phi$ at the minimum of $V(\phi)$ is given by $\phi_m=m_\phi/\lambda^{1/2}$ (retaining only the positive value),  and the resulting equation of motion for $\phi$ can be written as
\begin{eqnarray}\label{eq:em_dump}
\ddot{\phi}+(3H+\Gamma)\dot{\phi}+m_\phi^2\phi-\lambda\phi^3=0.
\end{eqnarray}

\begin{figure}
		\centering
	    \includegraphics[scale=0.33]{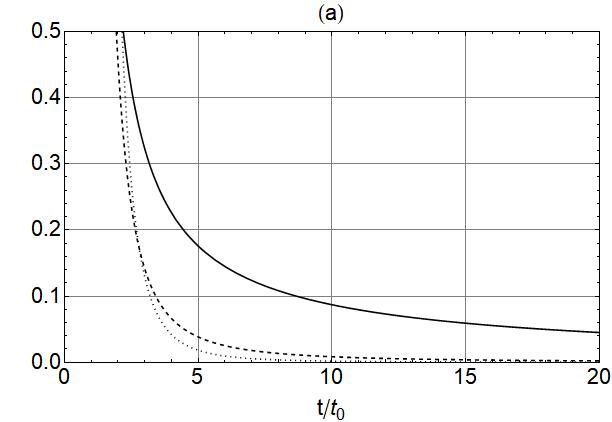}
		\includegraphics[scale=0.33]{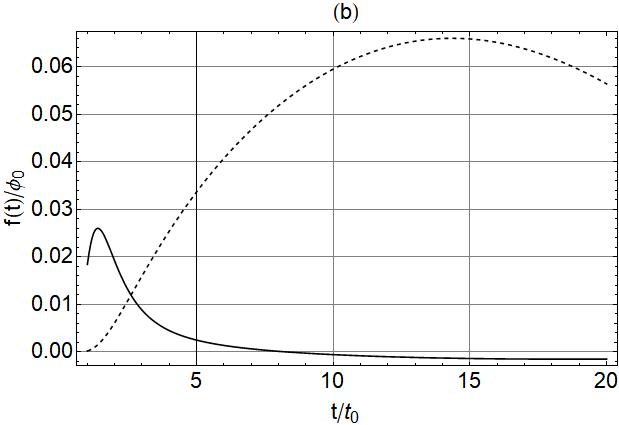}
		\caption{\label{fig:fig_hubble} Panel a) shows the evolution of $Ht_0$ (solid line), $|\dot{H}t_0^2|$ (dashed line) and $\ddot{H}t_0^3$ (dotted line), for $H$ given by (\ref{eq:hubble_par}). Panel b) shows the "residue" $r(t)/\phi_m$ (solid line) and the approximate solution $\phi/\phi_m$ (dashed line) given by (\ref{eq:solution_1}).}
	\end{figure}

The presence of the damping term $(3H+\Gamma)\dot{\phi}$ add some complications to the solution of \ref{eq:em_dump}. In order to propose a manageable solution for equation (\ref{eq:em_dump}), one first assumes the Hubble parameter during the inflation is given by \cite{J.Romero.M.Bellini.Nuovo Cim.B124.861-868.2009}
\begin{eqnarray}\label{eq:hubble_par}
    H=\frac{t_0}{t^2}\left[1-\ln\left(\frac{t}{t_0}\right)\right]+\frac{1}{t},
\end{eqnarray}

\noindent where $t\geq t_0$, being $t_0$ the time when inflation starts. Figure \ref{fig:fig_hubble}a) shows the evolution of $Ht_0$, $|\dot{H}t_0^2|$ and $\ddot{H}t_0^3$ for $t\geq t_0$. It is possible to see that for $t_0\leq t\lesssim 5t_0$, the Hubble parameter given by (\ref{eq:hubble_par}) fastly decreases, becoming almost flat for $5t_0<t$. Therefore, in some situations, the time-dependent behavior of $H$ can be neglected.

Analytic solutions for the case where one can neglect the damping term $(3H+\Gamma)\dot{\phi}$ are well-known, involving Jacobi elliptic functions. For our purpose, one supposes the approximate solution for equation (\ref{eq:em_dump}) can be written as
\begin{eqnarray}\label{eq:solution_11}
    \phi(t)\approx \phi_m e^{-(3H+\Gamma) t}\sin(m_\phi t),
\end{eqnarray}

\noindent which depends strongly on the behavior of $H$. Replacing the approximate solution (\ref{eq:solution_11}) into differential equation (\ref{eq:em_dump}), then it furnishes the "residue"
\begin{eqnarray}
\begin{split}
    r(t) & =e^{-3 t (3H+\Gamma)} [\lambda\phi_m^3 \sin^3(m_\phi t)  +  e^{2 t (3H +\Gamma)}
    \phi_m (-m_\phi \cos(m_\phi t) (3H+\Gamma  + 6 t \dot{H})  + \\
     & + \sin(m_\phi t) (-2 m_\phi^2 + 
         3\dot{H} (-2 + \Gamma t + 3 t (H + t\dot{H})) - 
        3 t \ddot{H}))].
\end{split}
\end{eqnarray}

The behavior of $r(t)/\phi_m$ is shown by the solid line in Figure \ref{fig:fig_hubble}(b) while the dashed line corresponds to the behavior of the approximate solution $\phi/\phi_m$. For both curves, one uses $m=0.1$ GeV, $\lambda=0.1$, $h=0.01$ GeV$^{1/2}$ and $\phi_e=0.1\phi_m$. As can be viewed, for $t_0\leq t< 5t_0$ the Hubble parameter dominates whereas for $t\gtrsim 5 t_0$ this parameter tends to vanish. Thus, as $5t_0\leq t$ grows, the ``residue" becomes less important, resulting the ansatz (\ref{eq:solution_11}) tend to be an \textit{asymptotic} solution for (\ref{eq:em_dump}) when the Hubble parameter turns to be almost constant.

To apply the Gouy-Stodola theorem, one should analyse the possible mean free path for $\phi$, which is not an easy task. One defines here $l$ as the distance traveled by $\phi$ without any collision or decay, allowing us to calculate the work due to the scalar field $\phi$.

The required knowledge of initial physical conditions of the inflationary epoch remains unknown, but it seems reasonable to suppose these conditions are close to the Quark-Gluon Plasma (QGP) than the classical plasma, for example. The QGP regime is characterized by a high-temperature regime, $T>\!\!>\Lambda_{QCD}$, where $\Lambda_{QCD}$ is the energy scale of Quantum Chromodynamics (QCD). Moreover, for this energy regime, the quarks and gluons are free (nonconfinement of QCD), and the running coupling constant of QCD is much smaller than 1, $\alpha_s=\alpha_s(T)<\!\!< 1$, where $Q_s\approx T$ at the saturation scale ($Q_s$ is the transferred momentum in the gluon frame). It is important to emphasize that the temperature in the inflationary epoch is, at least, of the order of the electroweak\new{?} epoch $T\approx 100\sim 200$ GeV \cite{B.Ryden.Introduction.to.Cosmology.Addison-Wesley.2003}, higher than the saturation scale.

The mean free path in such a system may depend on the energy available, the particle mass, the system temperature, the running coupling constant of QCD, among others parameters (density medium and volume, for example). The viscosity coefficients \cite{C.W.Misner.Astrophys.J.151.431.1968,O.Gron.Astrophys.Space.Science.173(2).191.1990} can also be considered to the free mean path calculation. However, in general, only the bulk viscosity is considered because the usual assumption of isotropy of the universe \cite{E.Komatsu.etal.WMAP.Collaboration.Astrophys.J.Suppl.192.18.2011,P.A.R.Ade.etal.Planck.Collaboration.Astron.Astrophys.594.A13.2016} and, consequently, the shear viscosity is neglected. The bulk viscosity, on the other hand, acts to enhance the entropy in the Friedmann-Robertson-Walker evolution \cite{R.Treciokas.G.F.R.Ellis.Comm.Math.Phys.23.1.1971,K.Sakai.Prog.Theo.Phys.46.1292.1971,M.Heller.M.Szydlowski.Astrophys.Space.Sci.90.327.1983}.

The mean free path is temperature-dependent and its functional form varies according to the constituents of the system. For example, the neutrino mean free path depends on $T^{-5}$ in the early universe, while in the limit $T\lesssim m_e$ where Klein-Nishima corrections are small and, the photon mean free path depends on $(n_e\sigma_T)^{-1}\sim T^{-1}$, where $n_e$ is the number density of electron and positrons in the medium as well $\sigma_T$ is the Thomson cross section. 

At the very beginning of inflation, the mean free path for $\phi$ may be comparable to the scale factor $a(t)$, implying the decay of $\phi$ starts when $l<a(t)$. Each oscillation of $\phi$ around the minimum of $V(\phi)$ is damped, reducing the mean free path due to the decaying rate of $\phi$ into $\chi$ particles. Thus, the number of particles in the medium is enhanced as $\phi$ oscillates and, consequently, $l$ diminishes. Observe the decaying rate $\Gamma$ is taken into account in the equation of motion for $\phi$. 

For a system composed of heavy quarks in the QGP regime, the mean free path depends on the energy loss and for $l$ smaller than the size of the system, one has \cite{S.Peigne.A.V.Smilga.Phys.Usp.52.659-685.2009}
\begin{eqnarray}
l\simeq \frac{\mu^2}{\alpha_s^2 T^3}
\end{eqnarray}

\noindent where $\mu$ is the Debye screening mass of the particle. The Debye screening mass depends on the proper definition of $\alpha_s$ in the nonconfinement phase of QCD, which is a subject far from agreement \cite{S.D.Campos.Int.J.Mod.Phys.A36.2150084.2021}. Recently, a mass generation mechanism was carried out for particles in QED  using the
Debye screening \cite{C.A.Bonin.G.B.deGarcia.A.A.Nogueira.B.M.Pimentel.Int.J.Mod.Phys.A35.2050179.2020}.

The energy of the scalar field $\phi$ is proportional to the square of wave amplitude, which allows us to assume the  following ansatz to the mean free path in the inflationary epoch
\begin{eqnarray}
l=\frac{\phi^2}{\alpha_s^2T_0^3},
\end{eqnarray}

\noindent where $l$ explicitly depends on the solutions for $\phi$, and $T_0$ is the initial temperature taken as constant during the process. In a more realistic situation, $T_0$ should be released; the inflaton decreases due to the damping term in (\ref{eq:em_dump}). Hence, at the end of the inflationary epoch, $l$ vanishes independently of the value attributed to $T_0$. In other words, the dumped solution (\ref{eq:solution_11}) is a consequence of both the decaying rate $\Gamma$ and the growth of $H$. Therefore, the number of $\phi$ particles vanishes as $T$ grows, also implying $l\rightarrow 0$ since there is no physical meaning in the existence of a mean free path for $\phi$ in a system where it does not exist anymore.  




By the nature of the scalar field $\phi$, which has an energy unit, the damping term in the equation of motion (\ref{eq:em_dump}) does not represent a true force since $\dot\phi$ posses a square energy unit, while $(3H+\Gamma)$ has an energy unit. To recover the square energy unit, one simply write the damping force as $-(3H+\Gamma)\dot{\phi}/m$, where $m=1$ GeV is a mass scale.

Then, taking into account all the above discussion, the irreversible work done by the damping force during the mean free path $l$ is given by
\begin{eqnarray}
W_{ir}=-\frac{l(3H+\Gamma)\dot{\phi}}{m_\phi} =-\frac{\phi^2}{m_\phi\alpha_s^2T_0^3}(3H+\Gamma) \dot{\phi},
\end{eqnarray}

\noindent which can be used to find out the entropy production using the Gouy-Stodola theorem 
\begin{eqnarray}\label{eq:entropia}
    \dot{\Sigma}=\frac{2 (3H+\Gamma) \dot{\phi^2}\phi + 
   \phi^2[3\dot{H}\dot{\phi} + (3H+\Gamma) 
        \ddot{\phi}]}{m_\phi\alpha_s^2 T_0^4}.
\end{eqnarray}

\begin{figure}
		\centering
	    \includegraphics[scale=0.30]{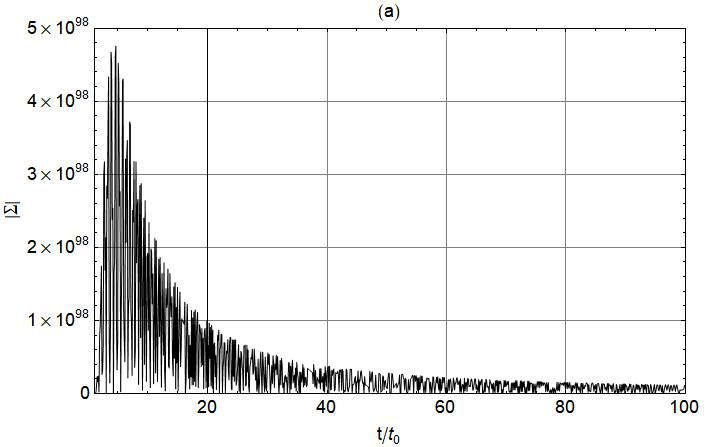}
		\includegraphics[scale=0.30]{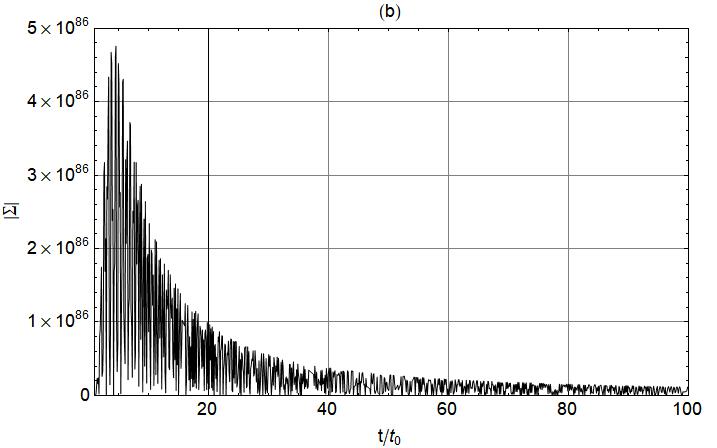}
        \includegraphics[scale=0.30]{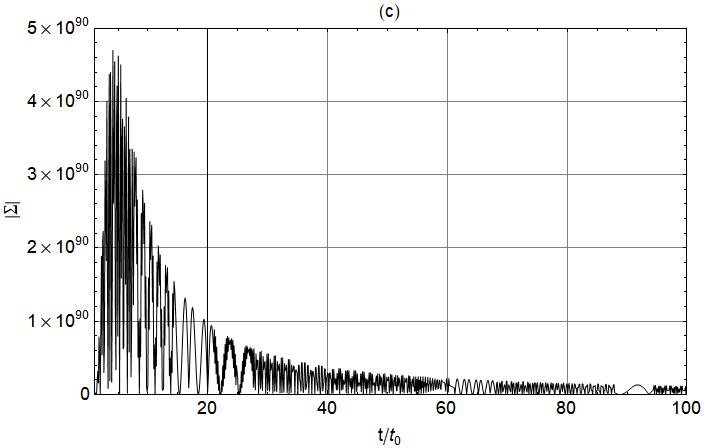}
		\includegraphics[scale=0.30]{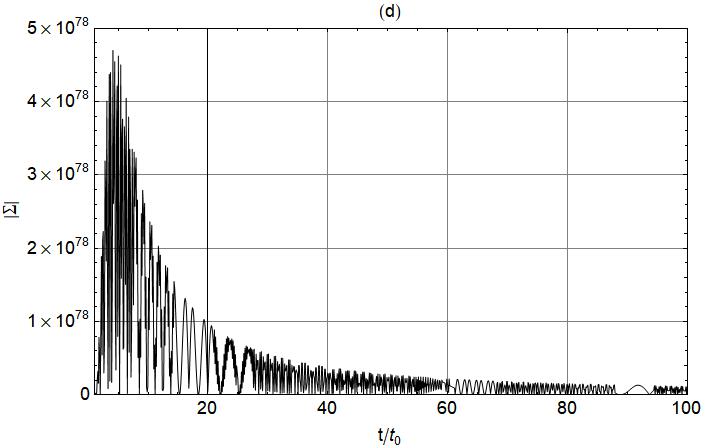}
		\caption{\label{fig:fig_entropia}Absolute value of entropy $\Sigma$ due to the damped scalar field $\phi$ during the inflationary epoch. Panel a) and b) $m_\phi=10^{14}$ GeV and $\lambda=10^{-9}$ and $10^{-7}$, respectively. Panel c) and d) $m_\phi=10^{12}$ GeV and $\lambda=10^{-9}$ and $10^{-7}$, respectively. In all cases, $T_0=200$ GeV, $t_0=1$ GeV$^{-1}$, and $\sigma=0.01$.} 
	\end{figure}
	
	\begin{figure}
		\centering
	    \includegraphics[scale=0.30]{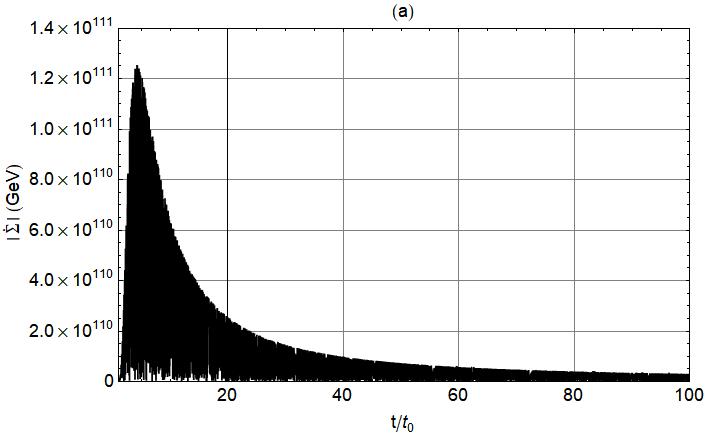}
		\includegraphics[scale=0.30]{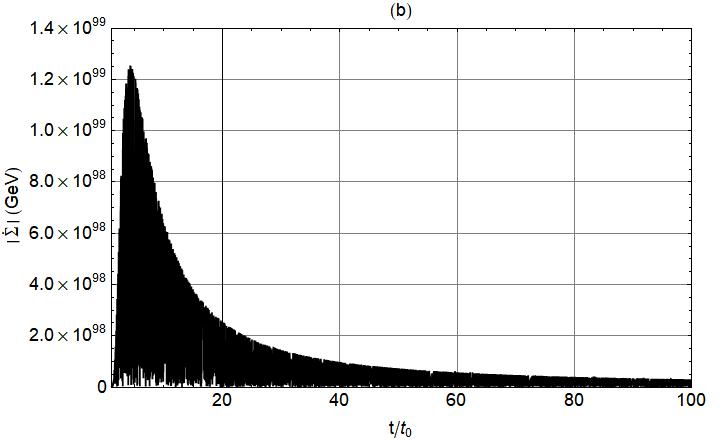}
		\includegraphics[scale=0.30]{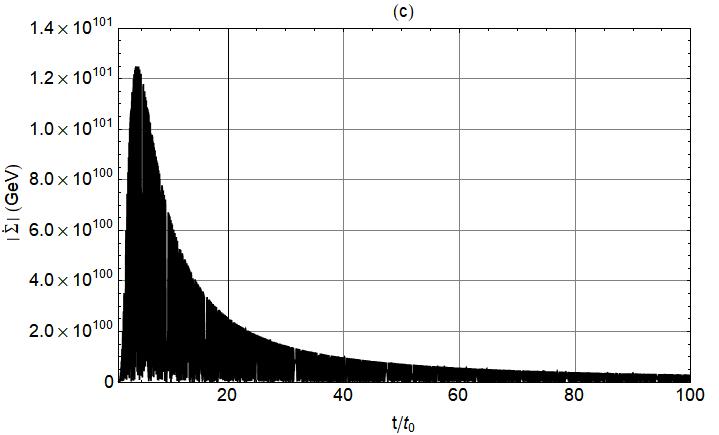}
		\includegraphics[scale=0.30]{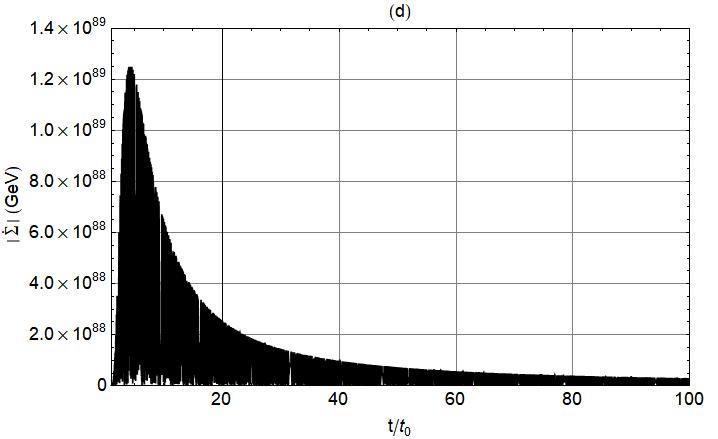}
		\caption{\label{fig:fig_entropiaprod}Absolute value of entropy production $\dot\Sigma$ due to the damped scalar field $\phi$ during the inflationary epoch. Panel a) and b) $m_\phi=10^{14}$ GeV and $\lambda=10^{-9}$ and $10^{-7}$, respectively. Panel c) and d) $m_\phi=10^{12}$ GeV and $\lambda=10^{-9}$ and $10^{-7}$, respectively. In all cases, $T_0=200$ GeV, $t_0=1$ GeV$^{-1}$, and $\sigma=0.01$.}
	\end{figure}

The inflaton mass ($m_{\phi}$) has been the subject of a long debate in the literature with a particular revival in recent years \cite{E.E.Basso.D.J.H.Chung.J.High.Energ.Phys.2021.146.2021,T.Moroi.W.Yin.J.High.Energ.Phys.2021.301.2021,D.Anselmi.J.Cosmol.Astropart.Phys.02.029.2021,M.Bastero-Gil.A.Berera.R.O.Ramos.J.G.Rosa.Phys.Lett.B813(10).136055.2021,M.P.Hertzberg.M.Jain.JCAP12.025.2020,Y.Aldabergenov.A.Chatrabhuti.S.V.Ketov.Eur.Phys.J.C79.713.2019,J.Haro.J.Amoros.S.Pan.Eur.Phys.J.C79(6).505.2019,G.Panotopoulos.N.Videla.Eur.Phys.J.C78.774.2018,A.Kaya.E.S.Kutluk.JCAP01.026.2015,J.Mielczarek.M.Kamionka.A.Kurek.M.Szydlowski.JCAP07.004.2010,A.Anisimov.Y.Bartocci.F.L.Bezrukov.Phys.Lett.B671.211-215.2009,E.J.Copeland.A.R.Liddle.D.H.Lyth.E.D.Stewart.D.Wands.Phys.Rev.D49.6410-6433.1994}. Depending on the approach performed, it can vary from $10^{-4}$ GeV \cite{J.G.Rosa.L.B.Ventura.Phys.Rev.Lett.122.161301.2019} up to $10^{14}$ GeV \cite{R.Brout.arXiv:gr-qc/0201060.}. The temperature in our calculations is set as $T_0=200$ GeV. Notice, however, the strong dependence of $\Sigma$ and $\dot{\Sigma}$ on $T_0$ may reduce drastically both values since upping in one order the temperature implies a decrease of four orders in $\Sigma$ and $\dot{\Sigma}$ (fine-tuning problem). 

According to the Planck Collaboration results \cite{P.A.R.Ade.etal.Planck.Collaboration.Astron.Astrophys.594.A20.2016}, the inflaton must follow a slow-roll trajectory in a plateau-like potential for large values of $\phi$. Then, the higher the mass of $\phi$, fewer the value of $\lambda$ to keep the flatness of $V(\phi)$ around its minimum. The value of $\lambda$ can vary according to the approach, of course. For the string theory approach called M-flation \cite{A.Ashoorioon.B.Fung.R.B.Mann.M.Oltean. M.M.Sheikh-Jabbari.JCAP.03.020.2014}, one can observe very small values, $\lambda\sim 10^{-14}$. Also, the gauge-invariant inflaton approach \cite{R.Allahverdi.B.Dutta.A.Mazumdar.Phys.Rev.Lett.99.261301.2007} has a very small self-coupling constant, $\lambda\sim 10^{-12}$. However, for the Higgs inflation model, where the mass for the decaying field is of the order of the top quark mass \cite{M.P.Hertzberg.M.Jain.JCAP12.025.2020}, the self-coupling has a large value, $\lambda\sim 10^{-2}\sim 10^{-1}$ \cite{M.P.Hertzberg.M.Jain.JCAP12.025.2020,W.Buchmuller.V.Domcke.K.Kamada.Phys.Lett.B726.467_470.2013,A.Anisimov.Y.Bartocci.F.L.Bezrukov.Phys.Lett.B671.211-215.2009}. Due to the necessary flatness of the potential and the large value expected for the entropy in the inflationary epoch, $\Sigma \sim 10^{88} \sim 10^{100}$ \cite{J.Romero.M.Bellini.Nuovo Cim.B124.861-868.2009,R.Brustein.A.J.M.MedvedPhys.Rev.D101.123502.2020}, one assumes small values for $\lambda$ and large for $m_\phi$.

Taking into account the above discussion, one assumes $h=0.01$ and $T_0=200$ GeV, resulting in the two free adjusting parameters, $m_\phi$ as well $\lambda$. Figure \ref{fig:fig_entropia} shows the entropy production given by result (\ref{eq:entropia}). As aforementioned, the functional form of the Hubble parameter depends on the specific features of the scale factor $a(t)$, leading to a strongly dependence in the entropy and entropy production in the choice of $a(t)$. Figure \ref{fig:fig_entropia} shows the entropy due to $\phi$: in panel a) and b), one uses $m_\phi=10^{14}$ GeV and $\lambda=10^{-9}$ and $\lambda=10^{-7}$, respectively. On the other hand, panel c) and d), one keeps $m_\phi=10^{12}$ GeV and uses $\lambda=10^{-9}$ and $10^{-7}$, respectively. One obtains a large entropy due to the inflaton decay during the inflationary epoch, where larger values are given by $\lambda=10^{-9}$ for both masses adopted for $\phi$. Of course, the accumulated entropy grows as $t/t_0$ rises.


Figure \ref{fig:fig_entropiaprod} shows the entropy production using the same parameters and the same sequence of results as for Figure \ref{fig:fig_entropia}. The large entropy production, in any case, occurs for $t<10t_0$, confirming the results shown in Figure \ref{fig:fig_entropia}. The accumulated entropy production rate also grows as $t$ rises. 

\section{Final Remarks}\label{sec:fr}

The laws of thermodynamics are cornerstones of physics. In principle, every physics system should be explained by using these laws. Of course, the complexity of some systems may prevent a thermodynamic treatment.

The Gouy-Stodola theorem may be a useful tool to evaluate the entropy production rate in irreversible systems by analyzing its thermodynamics. As a toy model, one uses this theorem to study the thermodynamics of a simple pendulum subject to a damping force.

In a more complex situation, one applies this theorem to study the entropy production rate in the inflationary epoch. In general, one expects huge values for the entropy in this regime, $\Sigma\approx 10^{88}$. In our calculations, high values for the entropy as well as the entropy production rate can be reached, showing that the Gouy-Stodola theorem may lead to similar results. 
	
	\section*{Acknowledge}
RHL and SDC thanks to UFSCar for the financial support.

\end{document}